\documentclass[aps,prl,twocolumn,showpacs,superscriptaddress,nofootinbib]{revtex4}  
\usepackage{graphicx}  
\usepackage{dcolumn}   
\usepackage{bm}        
\usepackage{braket}
\usepackage{exscale}                  
\usepackage[intlimits]{amsmath}       
\usepackage{amsfonts}
\usepackage{amssymb,amscd}
\usepackage{color}
\usepackage{hyperref}
\usepackage{subfigure}
\usepackage[normalem]{ulem}

\begin{document}

\title{Anomaly-induced dynamical refringence in strong-field QED} 

\author{N. Mueller}
\email{n.mueller@thphys.uni-heidelberg.de} 
\affiliation{Institut f\"{u}r Theoretische Physik, Universit\"{a}t Heidelberg, 69120 Heidelberg, Germany}
\author{F. Hebenstreit}
\email{hebenstreit@itp.unibe.ch}
\affiliation{Albert Einstein Center, Institut f\"{u}r Theoretische Physik, Universit\"{a}t Bern, 3012 Bern, Switzerland}
\author{J. Berges}
\email{j.berges@thphys.uni-heidelberg.de}
\affiliation{Institut f\"{u}r Theoretische Physik, Universit\"{a}t Heidelberg, 69120 Heidelberg, Germany}

\begin{abstract}
We investigate the impact of the Adler-Bell-Jackiw anomaly on the 
nonequilibrium evolution of strong-field quantum electrodynamics (QED) using real-time lattice gauge theory techniques. For field strengths exceeding the Schwinger limit for pair production, we encounter a highly absorptive
medium with anomaly-induced dynamical refractive properties. In contrast to earlier expectations based on equilibrium properties, where net anomalous effects vanish because of the trivial vacuum structure, we find that out-of-equilibrium conditions can have dramatic consequences for the presence of quantum currents with distinctive macroscopic signatures. We observe an intriguing tracking behavior, where the system spends longest times near collinear field configurations with maximum anomalous current. Apart from the potential relevance of our findings for future laser experiments, similar phenomena related to the chiral magnetic effect are expected to play an important role for strong QED fields during initial stages of heavy-ion collision experiments.  
\end{abstract}

\pacs{12.20.Ds, 
      11.15.Ha, 
      11.30.-j} 
   
\maketitle

{\it Introduction.} 
Quantum phenomena play an important role for the interactions of light with matter described by quantum electrodynamics (QED). A most spectacular quantum process is electron-positron pair production from the vacuum in the presence of very strong electric fields exceeding the Schwinger limit of about $10^{16}\,\text{V}/\text{cm}$~\cite{Sauter:1931zz,Heisenberg:1935qt,Schwinger:1951nm}. Recent advances in laser technologies  bring the observation of such non-perturbative dynamical phenomena into reach~\cite{Marklund:2008gj,DiPiazza:2011tq}, and possible experimental signatures have been investigated extensively~\cite{Blaschke:2005hs,Hebenstreit:2009km,Heinzl:2010vg,Kohlfurst:2013ura,Hebenstreit:2014lra,Otto:2014ssa}.

Remarkably, one of the hallmarks of quantum field theory -- the breaking of classical symmetries by radiative quantum corrections or so-called quantum anomalies -- has not played an important role in this context yet. This is partly a consequence of the very demanding computational effort: In QED, the Adler-Bell-Jackiw axial anomaly~\cite{Adler:1969gk,Bell:1969ts} involves the presence of a magnetic field and the relevant nonequilibrium quantum mechanical time evolution of the system requires a fully 3+1 space-time dimensional treatment. However, out-of-equilibrium conditions can have dramatic consequences for anomaly-induced electric currents: While in Abelian gauge theories such as QED it is well known that net anomalous effects can vanish in vacuum or thermal equilibrium because of the trivial vacuum structure, such a cancellation does not occur for far-from-equilibrium states encountered during real pair-production processes. 

In this Letter we demonstrate that for large electromagnetic fields the interplay between Schwinger pair production and the axial anomaly leads to a highly absorptive medium, whose anomalous refractive properties are dynamically induced by the produced particles and their non-linear interactions with the applied fields. We determine macroscopically observable consequences, such as plasma oscillations with an anomalous rotation of the electric field direction, which give direct information about the underlying quantum phenomena. This anomaly-induced dynamical refringence is fundamentally different from the dispersive phenomena that have been previously discussed, such as vacuum birefringence in a magnetic field~\cite{Klein:1964,Heinzl:2006xc}.        

Our results are obtained from ab initio calculations using
real-time simulation techniques for lattice QED with Wilson fermions following Refs.~\cite{Aarts:1998td,Berges:2010zv,Saffin:2011kc,Hebenstreit:2013qxa,Kasper:2014uaa,Buividovich:2015jfa,Gelfand:2016prm,Tanji:2016dka,Mueller:2016ven}. In this nonperturbative approach the full quantum nature of fermions is taken into account while the bosonic gauge-field dynamics is accurately represented by classical-statistical simulations for relevant field strengths. For the first time, this allows us to determine the nonequilibrium time evolution of QED in the presence of both magnetic as well as electric fields exceeding the Schwinger limit for pair production, including the crucial back-reaction of the induced currents on the applied fields. The necessary 3+1 dimensional description of the non-linear interplay between strong fields and produced matter becomes possible because of algorithmic advances \cite{Mueller:2016ven}, using large-scale computational resources~\cite{Comp}.    

Apart from the potential relevance of our findings for future laser experiments, similar phenomena are expected to play an important role for the strong QED fields appearing during the early stages of non-central collisions with relativistic heavy nuclei at the \textit{Large Hadron Collider} at CERN, or the \textit{Relativistic Heavy Ion Collider} at Brookhaven National Laboratory. The non-Abelian version of the axial anomaly in quantum chromodynamics (QCD) has attracted great interest in the context of the so-called chiral magnetic effect~\cite{Kharzeev:2007jp,Fukushima:2008xe}, which has been proposed as an explanation for the electric charge asymmetry in non-central heavy-ion collisions~\cite{Abelev:2009ac}.
Recently, the chiral magnetic effect has also been studied in condensed matter physics and its observation in Dirac semimetals has been reported \cite{Li:2014bha,Xiong:2015}. 

{\it Axial anomaly in QED.} 
Anomalies have the important consequence that certain constants of motion of the classical theory are no longer conserved when quantum effects are taken into account. In QED, the axial anomaly affects the axial current expectation value $j^\mu_5 = \braket{\bar{\psi}\gamma^\mu \gamma_5\psi}$ involving the different chiral components of the Dirac fermion field $\psi$ with Dirac matrices  $\gamma^\mu$ for $\mu = 0,1,2,3$ and $\gamma_5 = i \gamma^0 \gamma^1 \gamma^2 \gamma^3$. The four-divergence of this current is a sum of two distinct terms:
\begin{equation}
 \partial_\mu j_5^\mu=2im\braket{\bar{\psi}\gamma_5\psi} - \frac{e^2}{8\pi^2}F_{\mu\nu}\tilde{F}^{\mu\nu} \, .
 \label{eq:axialanomaly}
\end{equation}
While the origin of the contribution proportional to the electron mass $m$ can be understood from the mixing of the different chiral field components in the presence of a mass, the second term reflects the anomaly present in the quantum theory~\cite{Adler:1969gk,Bell:1969ts}. The anomalous contribution is described in terms of the electromagnetic field strength tensor $F_{\mu\nu}=\partial_\mu A_\nu-\partial_\nu A_\mu$ and its dual $\tilde{F}^{\mu\nu}=\frac{1}{2}\epsilon^{\mu\nu\rho\sigma}F_{\rho\sigma}$. In the absence of the anomalous term, Eq.~(\ref{eq:axialanomaly}) represents a continuity equation in the massless limit, for which the corresponding axial charge $Q_5=\int{d^3x j_5^0}=\int{d^3x\braket{\bar{\psi}\gamma^0\gamma_5\psi}}$ is classically conserved.

Moreover, by introducing the Chern-Simons current $K^\mu=4\epsilon^{\mu\nu\rho\sigma}A_\nu\partial_\rho A_\sigma$, the anomalous term in \eqref{eq:axialanomaly} can be written as a total divergence
\begin{equation}
 \partial_\mu K^\mu = F_{\mu\nu}\tilde{F}^{\mu\nu}=4\mathbf{E}\cdot\mathbf{B} \, ,
\end{equation}
which is related to the scalar product of the electric ($\mathbf{E}$) and the magnetic ($\mathbf{B}$) field. 
The total divergence reduces to boundary contributions upon space-time integration. Since the vacuum or thermal equilibrium state is trivial in QED -- unlike in non-Abelian gauge theories -- the relevant net production of axial charges in QED is vanishing for the time translation invariant equilibrium situations. In contrast, the non-trivial vacuum structure of non-Abelian gauge theories such as QCD leads to a non-conservation of the axial charge in this case~\cite{'tHooft:1976up,Christ:1979zm}. 

In the following, we point out that even for an Abelian gauge theory, such as QED, out-of-equilibrium conditions can have dramatic consequences for anomaly-induced currents. More specifically, we demonstrate that Schwinger pair production in strong-field QED leads to nonequilibrium states for which the axial anomaly results in a very significant axial charge density with intriguing observable consequences.   

{\it Nonequilibrium strong-field QED.}
In a realistic situation, the phenomenon of real electron-positron pair production represents a nonequilibrium problem for which a strong electromagnetic field is present at some initial time $t_i$. Once matter is produced, it will back-react on the applied field such that the total energy is conserved. In general, this situation is not time translation invariant and involves far-from-equilibrium states. The net axial charge produced at some later time $t_f > t_i$ is given by $Q_5(t_f) - Q_5(t_i)$, where typically neither $t_i$ nor $t_f$ may be taken to the remote past or infinite future. 

This represents a demanding initial-value problem in nonequilibrium quantum field theory, which can be treated from first principles employing functional integral techniques on the Schwinger-Keldysh closed time path~\cite{Schwinger:1960qe,Keldysh:1964ud}. An important simplification occurs in the strong-field regime, where the gauge field dynamics can be accurately mapped onto a classical-statistical problem which can be solved on a computer~\cite{Aarts:1998td,Berges:2010zv,Saffin:2011kc,Hebenstreit:2013qxa,Kasper:2014uaa,Buividovich:2015jfa,Gelfand:2016prm,Tanji:2016dka,Mueller:2016ven}. Since the fermion fields appear quadratically in the QED action, they may be taken into account without further approximations by using a mode-function expansion~\cite{Aarts:1998td} and discretizing the system on a hypercubic lattice using Wilson fermions following Refs.~\cite{Hebenstreit:2013qxa,Kasper:2014uaa}. The Wilson regularization ensures that the axial anomaly is correctly reproduced in the continuum limit \cite{Tanji:2016dka,Mueller:2016ven,Karsten:1980wd,Rothe:1998ba}.

\begin{figure}[t]
\centering
 \includegraphics[width=\columnwidth]{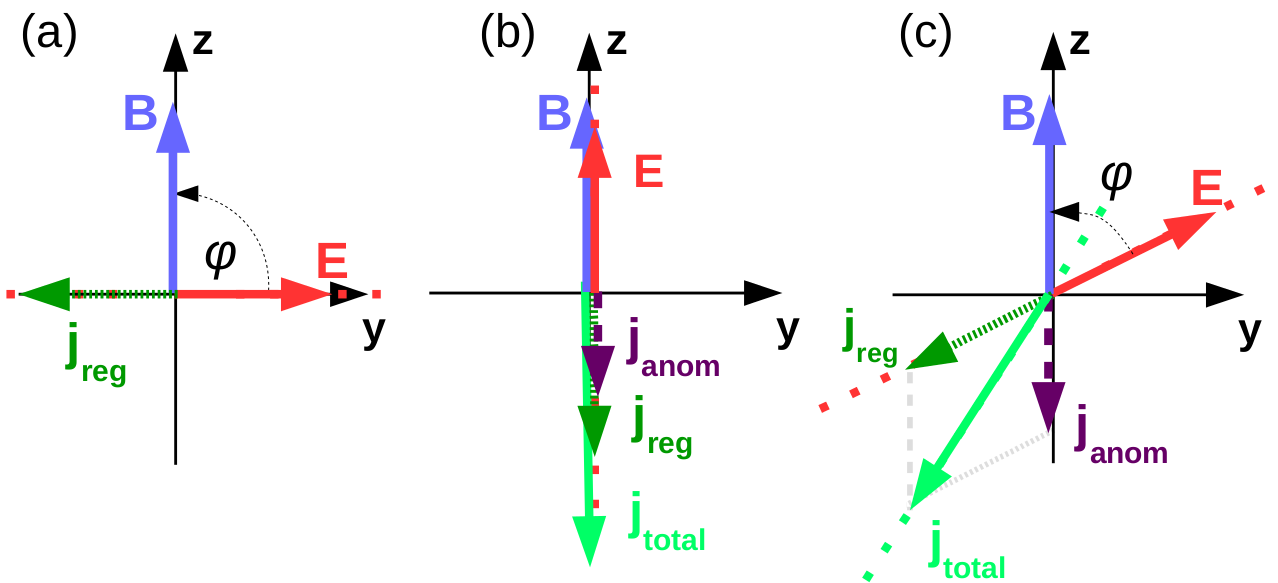}
 \caption{
Schematic current generation for different values of the angle $\varphi$ between the magnetic and electric field directions.
 (\textit{a}) For $\varphi=\pi/2$, only a regular current $j_\text{reg}$ is generated parallel to the electric field. 
 (\textit{b}) For $\varphi=0$, an anomalous current $j_\text{anom}$ is induced in addition to $j_\text{reg}$. 
 Both of them point in the same direction.
 (\textit{c}) For arbitrary values of the angle, $j_\text{reg}$ and $j_\text{anom}$ point in different directions.
 Accordingly, the total current $j_\text{tot}$ is not parallel to the electric field direction.}
 \label{fig1:current}
\end{figure}

For our studies, we choose initial conditions which correspond to the fermion vacuum, i.e.\ vanishing axial and vector charges, in the presence of a static magnetic field in the $z$-direction: $\mathbf{B}=B_0\mathbf{e}_z$.
In the fermion sector, this is implemented via diagonalization of the corresponding Dirac operator at initial time $t_i=0$.
We do not consider possible initial-state fluctuations in the gauge field sector as they play a minor role for the dynamics in the strong-field regime. Instead of initializing a static electric field $\mathbf{E}_0=E_0\mathbf{e}_\varphi$, where $\mathbf{e}_\varphi$ is the unit vector in the $y$--$z$--plane with angle $\varphi=\measuredangle(\mathbf{e}_z,\mathbf{e}_\varphi)$ (cf.~Fig.~\ref{fig1:current}), we trigger the dynamics by an external field pulse $\mathbf{E}_0=E_0(t)\mathbf{e}_\varphi$ with
\begin{equation}
 E_0(t)= E_{0}\, \text{sech}^2({\omega (t-t_0)})\, .
 \label{eq:pulse}
\end{equation}
Apart from the fact that this choice allows comparisons to earlier numerical studies with purely electric initial configurations~\cite{Hebenstreit:2011wk,Hebenstreit:2013qxa}, the pulse~\eqref{eq:pulse} is very efficient in producing axial and vector charges during a short time interval while not accelerating them too much, which reduces the computational cost and renders our numerical simulations manageable. We emphasize, however, that the qualitative behavior for the employed initial configuration and a static applied field is very similar. 

To accurately resolve low-momentum fermion and axial charge production, a large spatial volume \mbox{$V=\prod_{i=1}^3 a_iN_i$} is required, where $a_i$ denotes the lattice spacing and $N_i$ the number of lattice sites in the $i$--direction.
On the other hand, the ultraviolet properties are probed during the time evolution owing to the fact that electric fields accelerate particles to high momenta, which can only be properly resolved by small values of $a_i$.
We perform our simulations on a $20\times 20\times 40$ grid with lattice spacings $a_x=a_y=0.08m^{-1}$ and $a_z=0.06m^{-1}$, which reflects the anisotropy of the initial field configuration. 
Translation invariance in the $z$-direction allows to perform a partial Fourier transformation to decrease the numerical cost.
In the following, we initialize an external electric pulse \eqref{eq:pulse} with $E_0=20E_c$, $\omega=1.2m$ and $t_0=2.5m^{-1}$ at an initial angle of $\varphi(0)=25^{\circ}$  in the presence of a static magnetic field $B_0=4.9B_c$, where  
\begin{align*}
E_c&=m^2/e\simeq1.3\cdot10^{16}\,\text{V}/\text{cm}, \\ B_c&=m^2/e\simeq 4.4\cdot{10}^{13}\,\text{G}.
\end{align*}
Here we follow particle physics conventions, setting the speed of light $c$ and reduced Planck constant $\hbar$ to unity. 

{\it Anomaly-induced dynamical refringence.}
Exceeding the critical field strength $E_c$, virtual electron-positron fluctuations are expected to be separated to become real particles~\cite{Sauter:1931zz,Heisenberg:1935qt,Schwinger:1951nm}. Though magnetic fields cannot create particles directly, they can have an influence on particle production, introducing discrete energy levels as has been pointed out for collinear electric and magnetic fields with $\varphi=\measuredangle(\mathbf{E},\mathbf{B})=0$~\cite{Daugherty:1976mg,Kim:2003qp,Tanji:2008ku,Warringa:2012bq}. 
In fact, the parallel case is special since the induced current from produced particles always parallels the applied electric field, as is the case also for $\mathbf{E}$ and $\mathbf{B}$ perpendicular to each other (cf.~Fig.~\ref{fig1:current}). Most important, in the absence of the axial anomaly the induced current and the electric field would be collinear for any other values of the angle between $\mathbf{E}$ and $\mathbf{B}$. It is only the anomaly term in (\ref{eq:axialanomaly}) that leads, in addition to a regular current $j_\text{reg}$ parallel to $\mathbf{E}$, to the generation of an anomalous current component $j_\text{anom}$ parallel to $\mathbf{B}$. Consequently, for general $\varphi=\measuredangle(\mathbf{E},\mathbf{B})$ the presence of the quantum anomaly manifests itself in a distinctive macroscopic property of the total induced current.

\begin{figure}[t]
\centering
 \includegraphics[width=\columnwidth,height=5cm]{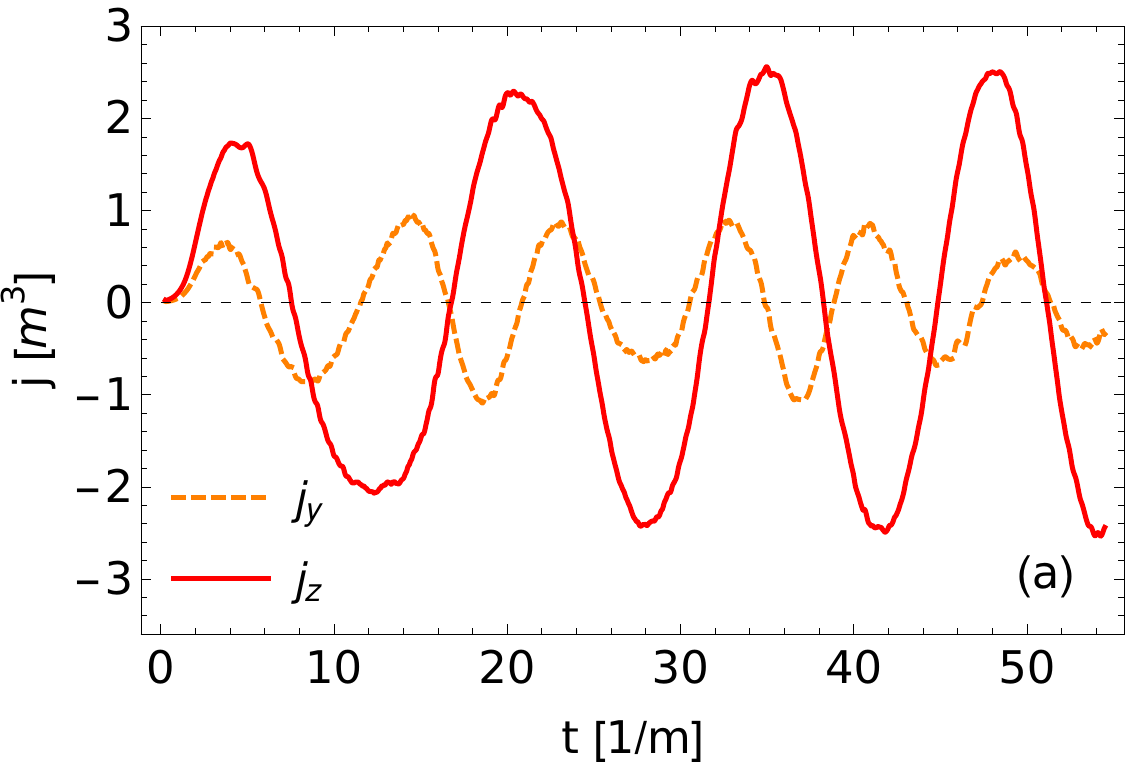}
 \includegraphics[width=\columnwidth,height=5cm]{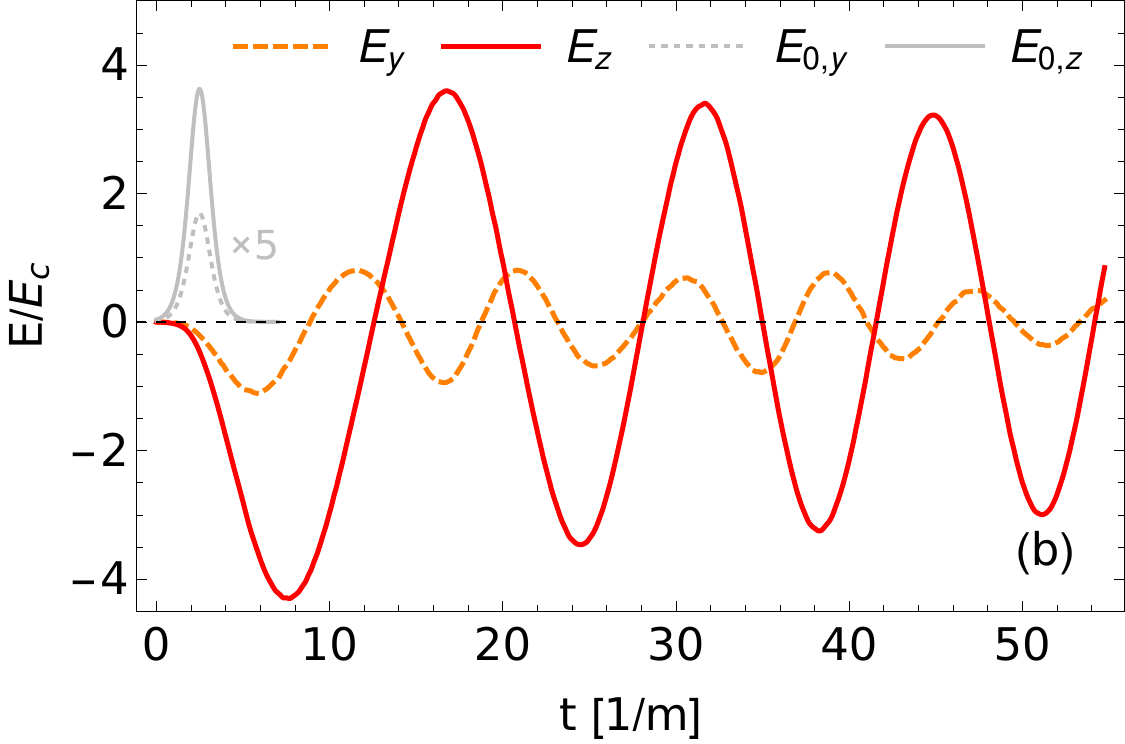}
 \caption{
Time-evolution of the non-vanishing electric current and electric field components.
 (\textit{a}) The total current results from a regular and an anomalous contribution. 
 Accordingly, the $y$-- and $z$--components oscillate out of phase.
 (\textit{b}) The $y$-- and $z$--components of the electric field are also out of phase. 
 As a consequence, $\mathbf{E}$ rotates relative to $\mathbf{B}$, cf.~Fig.~\ref{fig3:angle}. Shown is also the initial pulse whose amplitude is rescaled by a factor of five for better visualization.}
 \label{fig2:curfield} 
\end{figure}

In Fig.~\ref{fig2:curfield}, we display the time evolution of the $y$-- and $z$--components of the electric currents $j^i=e\braket{\bar{\psi}\gamma^i\psi}$ as well as the corresponding electric field components as a function of time. Since $\nabla\times\mathbf{E}(t)=0$, the magnetic field of the system remains constant.
Similar to a purely electric configuration, the interplay between particle production and subsequent screening of the electric field results in plasma oscillations~\cite{Kasper:2014uaa,Hebenstreit:2013qxa,Kluger:1992gb}. By comparing the upper and the lower graph of Fig.~\ref{fig2:curfield}, the zero crossings of each current component are seen to occur around the same time as the extremum of the respective electric field component and vice versa. However, in contrast to the Schwinger effect in a purely electric configuration, for which the electric current is always anti-parallel to the electric field so that all field components oscillate with the same plasma frequency, we now observe that the oscillations of the $y$-- and $z$--components are out of phase.

This intriguing behavior can be traced back to the production of an anomalous electric current along the magnetic field direction owing to the chiral magnetic effect~\cite{Kharzeev:2007jp,Fukushima:2008xe,Fukushima:2015tza}: The spin of particles tends to align in magnetic fields and results in a correlation between electric charge, chirality and momentum. Positively charged, right-handed particles move along the magnetic field lines while negatively charged right-handed particles move in the opposite direction. 
This effect is usually canceled by the inverse behavior of the left-handed particles.
For $\mathbf{E}\cdot\mathbf{B}\neq0$, however, an imbalance between right- and left-handed particles is produced such that an anomalous electric current along the magnetic field direction is generated,
\begin{equation}
\vec{j}_{anom}=\sigma_5(q_5(t))\vec{B},
\end{equation}
where $\sigma_5(q_5(t))$ is the dynamical anomalous conductivity, which cannot be reduced to a static quantity in our case.

\begin{figure}[t]
\centering
 \includegraphics[width=\columnwidth,height=5cm]{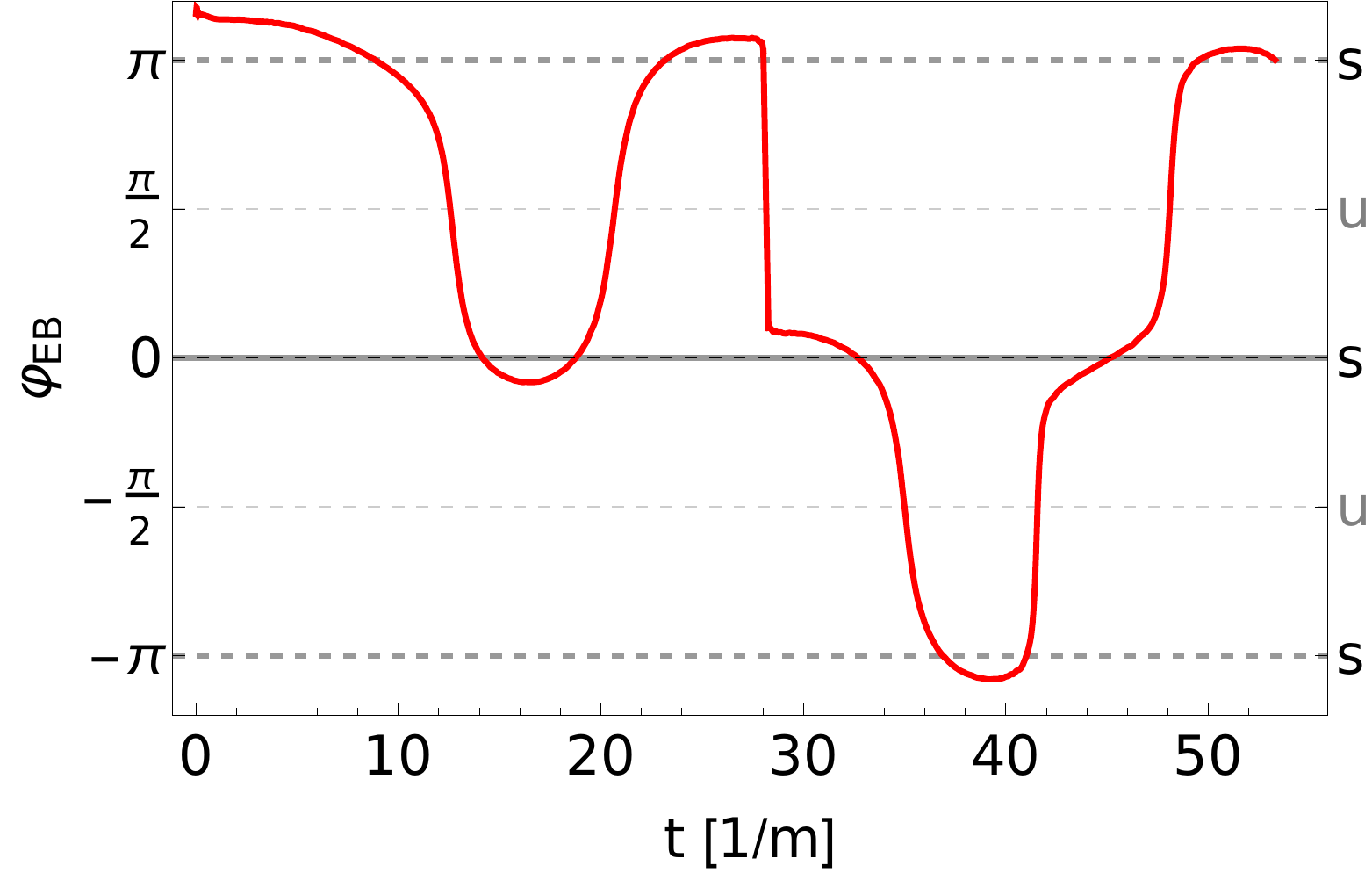}
 \caption{
Anomalous rotation of the dynamical electric field as described by the time-dependent angle \mbox{$\varphi(t)=\measuredangle(\mathbf{E}(t),\mathbf{B})$}. The evolution exhibits tracking behavior, where the system spends longest times near collinear field configurations with maximum quantum current.}
 \label{fig3:angle}
\end{figure}

We point out that the anomalous rotation of the electric field direction is in general an important consequence of this effect, and we display the time evolution of the angle $\varphi(t)$ in Fig.~\ref{fig3:angle}. Only for the specific initial values $\varphi(0)=0,\pi$ (parallel) and $\varphi(0)=\pm\pi/2$ (perpendicular) no anomalous rotation occurs (cf.~Fig.~\ref{fig1:current}). Most remarkably, we find that
for general initial configurations the evolution exhibits tracking behavior: Irrespective of the initial condition details, the system spends longest times near collinear field configurations with maximum quantum current, while disfavoring orthogonal fields with no anomalous current generation. In fact, $\varphi=\pm\pi/2$ turn out to be unstable stationary points ($u$), where even arbitrarily small deviations lead to the generation of axial charges and, accordingly, to an anomalous rotation of the electric field direction. The collinear tracking solutions (s) represent an important self-focusing mechanism, which makes our phenomenon of anomaly-induced dynamical refringence very robust and different from conventional dispersive phenomena described in terms of material constants.

\begin{figure}[t]
\centering
 \includegraphics[width=\columnwidth]{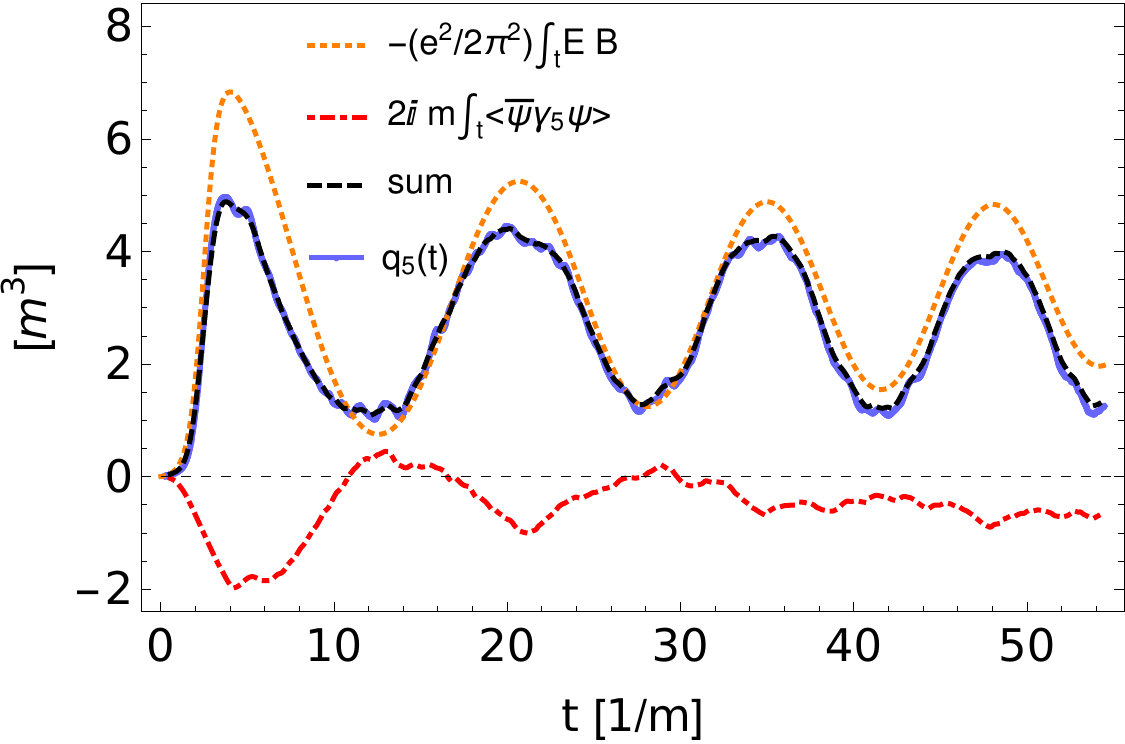}
 \caption{
Generation of the axial charge density $q_5$ as a function of time. In addition, the different contributions on the r.h.s.\ of the space-time integrated anomaly equation \eqref{eq:anomalyint} are shown. 
We note that in supercritical fields (as present in this case, cf.~Fig.\ref{fig2:curfield}) the recreation of axial charge via particle production dominates over the damping due to the finite fermion mass\cite{Grabowska:2014efa,Guo:2016nnq}, which is marked by the presence of the pseudoscalar condensate.%
}
 \label{fig4:anomaly}
\end{figure}

{\it Axial charge production.}  
Our simulations allow us for the first time to check the Abelian anomaly equation (\ref{eq:axialanomaly}) out of equilibrium by explicitly computing all the different terms appearing on its left and right hand side. For the time evolution of the axial anomaly in non-Abelian gauge theories we refer to Refs.~\cite{Saffin:2011kn,Tanji:2016dka}. We study the time-integrated and volume-averaged anomaly equation:
\begin{equation}
 q_5(t) = 2im \, \mkern-8mu\int_{0}^{t}{\braket{\bar{\psi}\gamma_5\psi}dt'} - \frac{e^2}{2\pi^2}\, \mkern-8mu\int_{0}^{t}{\mathbf{E}\cdot\mathbf{B}\,dt'} \, .
 \label{eq:anomalyint} 
\end{equation}
We first compute the axial charge density $q_5(t)$ directly. The solid (blue) line of Fig.~\ref{fig4:anomaly} clearly demonstrates the generation of the axial charge density and its nonequilibrium time evolution. In addition, we display the individual contributions appearing on the r.h.s.\ of \eqref{eq:anomalyint} with the time-integrated pseudoscalar condensate (first term/dotted orange curve) and anomaly term (second term/dashed-dotted red curve). The dashed (black) curve represents their sum and the agreement with $q_5$ within numerical errors reflects the underlying quantum anomaly and the ability of our methods to capture the intriguing phenomena associated to it. In fact, the exact reproduction of the anomaly equation is only expected in the continuum limit \cite{Karsten:1980wd,Rothe:1998ba}. 
While lattice artifacts from the Wilson term are negligible, the dominant error in our numerical results is due to the finite system size $V$ and lattice spacings $a_i$.
Based on restricted variations thereof, we estimate the total numerical error for the quantities considered to be of the order of a few percent, in overall agreement with the above consistency check of the anomaly equation.

{\it Conclusions.}
We have shown that the Adler-Bell-Jackiw anomaly, which does not
contribute in vacuum-to-vacuum transitions in QED, has dramatic consequences 
in strong-field QED out of equilibrium. Using ab initio real-time lattice techniques,
we demonstrated that axial charges can be produced during the time evolution of the unstable QED vacuum in the presence of nonperturbatively large electromagnetic fields. 
For general angles between the initial electric and magnetic fields, we 
discovered a highly absorptive medium, whose anomalous refractive properties 
are dynamically determined by tracking solutions with maximum quantum
current. 


In principle, this new medium is accessible to future high-intensity laser experiments where, e.g., two counter-propagating optical laser pulses produce a slowly varying, standing-wave magnetic field which is superimposed by a single attosecond pulse in the focal region~\cite{Sansone:2006}. The intriguing dynamics of Fig.~\ref{fig2:curfield} leads to an anomalous rotation of the fermionic current which is detectable by spectrometric measurements. Similar aspects of the chiral magnetic effect may also be studied during the early stages of non-central heavy-ion collisions, where electromagnetic probes such as direct photons represent important messengers of the nonequilibrium early-time dynamics, or in Dirac semimetals, where the underlying chiral magnetic effect has been observed \cite{Li:2014bha,Xiong:2015}. Related setups using ultracold quantum gases~\cite{Kasper:2015cca} may provide another very promising route to uncover the fundamental quantum phenomena that are predicted here. 

{\it Acknowledgments.} We thank V.~Kasper, L.~F.~Palhares, J.~Pawlowski, S.~Schlichting, S.~Sharma, N.~Tanji and R.~Venugopalan for discussions.
NM acknowledges support by the Studienstiftung des Deutschen Volkes and thanks Brookhaven National Laboratory, where this work was completed.
FH acknowledges support from the Alexander von Humboldt Foundation in the early stages of this work as well as from the European Research Council under the European Union's Seventh Framework Programme (FP7/2007-2013)/ ERC grant agreement 339220. This work is part of and supported by the DFG Collaborative Research Centre SFB 1215 (ISOQUANT).

\end{document}